\documentclass[twocolumn,superscriptaddress,preprintnumbers,amsmath,amssymb,prc]{revtex4-1}

\usepackage{graphicx}
\usepackage{dcolumn}
\usepackage{bm}
\usepackage{ifpdf}
\usepackage{epstopdf}
\usepackage{slashed}
\usepackage{amsfonts}
\usepackage{mathrsfs}
\usepackage{amssymb}
\usepackage{multirow}
\usepackage{CJK}
\usepackage{xcolor}
\usepackage{color}
\usepackage{ulem}

\begin{document}
\begin{CJK}{UTF8}{ipxm}
\preprint{RIKEN-iTHEMS-Report-21}

\title{Cooper quartet correlations in infinite symmetric nuclear matter}

\author{Yixin Guo (郭一昕)}
\email{guoyixin1997@g.ecc.u-tokyo.ac.jp}
\affiliation{Department of Physics, Graduate School of Science, The University of Tokyo, Tokyo 113-0033, Japan}
\affiliation{RIKEN iTHEMS, Wako 351-0198, Japan}

\author{Hiroyuki Tajima (田島裕之)}
\email{htajima@g.ecc.u-tokyo.ac.jp}
\affiliation{Department of Physics, Graduate School of Science, The University of Tokyo, Tokyo 113-0033, Japan}

\author{Haozhao Liang (梁豪兆)}
\email{haozhao.liang@phys.s.u-tokyo.ac.jp}
\affiliation{Department of Physics, Graduate School of Science, The University of Tokyo, Tokyo 113-0033, Japan}
\affiliation{RIKEN Nishina Center, Wako 351-0198, Japan}

\date{\today}

\begin{abstract}
We investigate the quartet correlations in four-component fermionic systems at the thermodynamic limit within a variational many-body theory.
The Bardeen-Cooper-Schrieffer (BCS)--type variational wave function is extended to the systems with the coexistence of pair and quartet correlations at zero temperature.
Special attention is paid to the application of the present framework to an $\alpha$-particle condensation in symmetric nuclear matter, where the coexistence of deuteron and $\alpha$ condensations is anticipated.
We also discuss how physical properties, such as  quasiparticle dispersion, can be modified by the pair and quartet correlations and show a hierarchical structure of in-medium cluster formations in infinite nuclear matter.
The present results may also contribute to the interdisciplinary understanding of fermionic condensations beyond the BCS paradigm in many-body systems.
\end{abstract}

\maketitle

\section{Introduction}

Study of quantum many-body phenomena is one of the central issues in modern physics.
A striking example is superconductivity, where two electrons form a Cooper pair due to the Fermi-surface instability and the many-body wave function develops macroscopic coherence as a result of the condensation of Cooper pairs.
Such a non-trivial phenomenon can be successfully described by the Bardeen-Cooper-Schrieffer (BCS) theory~\cite{BCS}, which also made significant impacts on various research fields such as nuclear and particle physics.

In pure neutron matter, neutron superfluidity with the isovector pairing, that is, the spin-singlet $^1S_0$ and spin-triplet $^3P_2$ Cooper pairs, has been widely discussed~\cite{Dean,Takatsuka}.
Such a superfluid state can be observed in neutron stars, where a cooling process of a star~\cite{Pethick} and the so-called glitch phenomena~\cite{Sauls} may involve nontrivial aspects of neutron superfluids.
Moreover, the deuteron condensation has also been anticipated in the symmetric nuclear matter due to strong neutron-proton interactions in the isoscalar channel~\cite{Baldo}.
In particular, the crossover from the Bose-Einstein condensation (BEC) of deuterons to the BCS-type neutron-proton pairing state driven by the increase of nucleon density has been investigated theoretically~\cite{Lombardo,Huang,Jin,Stein,Stein2}.
Such a phenomenon is called BCS-BEC crossover, which has been realized in ultracold Fermi atomic gases~\cite{Regal,Zwierlein,Kinast} and in superconductors~\cite{Shibauchi,Nakagawa}, and its connections to nuclear systems have also been studied extensively~\cite{Strinati,Ohashi}.
In asymmetric nuclear matter, both the isovector and isoscalar pairing states can appear and compete with each other~\cite{Mao,Tajima,Yan}.
While the isoscalar pairing interaction is stronger than the isovector one, the imbalance between neutron and proton Fermi levels suppresses the isoscalar neutron-proton pairing.
In this regard, a variety of pair-condensation phases such as Fulde-Ferrel-Larkin-Ovchinikov-like state~\cite{FF,LO} have been proposed in this system~\cite{Strinati}.

On the other hand, because the nuclear matter generally consists of four kinds of nucleons with spin and isospin degrees of freedom, the four-body (called quartet) correlations will play a significant role.
Indeed, an $\alpha$ particle, which is a spin- and isospin-singlet bound state consisting of two neutrons and two protons, is known as one of the most stable nuclear clusters due to its large binding energy $E_\alpha=28.29$~MeV.
Accordingly, the investigation on four-body clusters composed of two neutrons and two protons coupled to the isospin $T = 0$ and to the angular momentum $J = 0$, which is commonly called an $\alpha$-like quartet, has become a long-standing issue.
Such a four-body structure has been investigated theoretically in finite nuclei~\cite{Bremond,Flowers,Senkov,Sandulescu,Sambataro,Baran,Baran2,Sambataro2}.
In addition, while the $\alpha$-particle condensation temperature~\cite{Ropke,Schuck},
the density-induced suppression of the $\alpha$-condensate fractions~\cite{Funaki,Sogo2010},
the polaronic $\alpha$ particles~\cite{Nakano,Moriya},
and the thermal four-body correlations~\cite{Ropke2}
have been examined theoretically in infinite nuclear matter,
it is worth exploring how the quartet correlation affects physical properties of infinite nuclear matter at zero temperature, which is relevant to neutron star physics.
The pair and quartet condensations are also important to clarify the microscopic origin of symmetry energy in dilute nuclear matter.

Moreover, the theoretical framework of quartet correlations in infinite matter
can be applied to other systems such as biexciton condensation~\cite{Chase,XXHuang}, SU(4) Fermi atomic gases~\cite{Wu,Ozawa,Miyazaki}, and charge $4e$ superconductors~\cite{Babaev2004, Berg, Herland2010, Fernandes}.
In the context of interdisciplinary studies of multi-component fermions,
more than two-body cluster states induced by the Cooper instability~\cite{Kamei,Tsuruta,Niemann,Kirk,Akagami,Tajima2,Tajima3} and ground-state properties and $N$-particle off-diagonal long-range order of $\eta$ pairing in the attractive SU($N$) Hubbard model~\cite{Yoshida,Yoshida2}
have been studied theoretically.
Also, the recent experiment indicates the existence of quartet correlations with broken time-reversal symmetry above the superconducting critical temperature in a condensed-matter system
\cite{Grinenko}.

In this study, we examine the quartet correlations in the infinite nuclear matter at the thermodynamic limit within the many-body variational approach.
The BCS-type trial wave function in the momentum space is extended to the symmetric nuclear matter, where the pair and quartet correlations are described by the coherent superposition of neutron-proton Cooper pairs and $\alpha$-like Cooper quartets.
We show an analytic structure of the quartet BCS wave function in terms of nucleon degrees of freedom and how Cooper pairs and quartets are formed in cold symmetric nuclear matter at the thermodynamic limit.
In particular, solving the variational equations for the quartet BCS framework, we present the numerical results of the quasiparticle dispersion and the variational parameters describing the momentum distribution of nucleons under the presence of deuteron and quartet correlations.

This paper is organized as follows.
The theoretical framework is presented in Sec.~\ref{sec:II}, where we show a variational formalism for pair and quartet condensations in symmetric nuclear matter.
In Sec.~\ref{sec:III}, we discuss how physical properties are modified by the coexistence of pair and quartet condensations in the present framework.
Finally, we summarize this paper in Sec.~\ref{sec:IV}.
In the following, we take $\hbar=c=k_{\rm B}=1$.

\section{Theoretical framework}\label{sec:II}

\subsection{Hamiltonian}\label{sec:IIA}

In the infinite nuclear matter, the isovector and isoscalar pairing correlations can be described by the Hamiltonian
\begin{align}
\label{hamiltonian}
H=\,&
\sum_{\bm{p}, s_z}  \left(
\varepsilon_{\nu, \bm{p}}
\nu_{\bm{p}, s_z}^{\dagger} \nu_{\bm{p}, s_z}
+\varepsilon_{\pi, \bm{p}}
\pi_{\bm{p}, s_z}^{\dagger} \pi_{\bm{p}, s_z} \right)\nonumber\\
&+\frac{1}{2}\sum_{\bm{P}, \bm{q}, \bm{q}'}
\sum_{T_3=-1}^{+1}
P_{1, T_3}^{\dagger}(\bm{P}, \bm{q})
V_s\left(\bm{q}, \bm{q}'\right)
P_{1, T_3}\left(\bm{P}, \bm{q}'\right)\nonumber\\
&+ \frac{1}{2} \sum_{\bm{P}, \bm{q}, \bm{q}'}
\sum_{S_z=-1}^{+1}
D_{1, S_z}^{\dagger}(\bm{P}, \bm{q})
V_t\left(\bm{q}, \bm{q}'\right)
D_{1, S_z}\left(\bm{P}, \bm{q}'\right),
\end{align}
where the creation operators $\nu^\dagger$ and $\pi^\dagger$ create a neutron and a proton, respectively;
$\bm{p}$ is the single-particle momentum, $\bm{q}=\frac{1}{2}(\bm{p}_1-\bm{p}_2)$ is the relative momentum, and $\bm{P}=\bm{p}_1+\bm{p}_2$ is the center-of-mass momentum;
$T_3$ is the third component of isospin (or $\tau_3$ for single nucleon), $S_z$ is the third component of spin (or $s_z$ for single nucleon);
and $V_t$ and $V_s$ are the interaction strengths in the isoscalar and isovector channels, respectively.
In addition, the single-particle energy reads
$\varepsilon_{i, \bm{p}}=
\frac{\bm{p}^{2}}{2 M_i}-\mu_{i}$ ($i = \pi, \nu$),
where $\mu_{i}$ is the nucleon chemical potential and $M_{i}$ is the nucleon mass.

The two-nucleon annihilation operators are defined as
\begin{subequations}
\begin{align}
P_{1, T_3}(\bm{P}, \bm{q})
=\,&
\sum_{s_z, s'_z}
\sum_{\tau_3, \tau'_3}
C_{\frac{1}{2} \frac{1}{2} s_z s'_z}^{00}
C_{\frac{1}{2} \frac{1}{2}\tau_3\tau'_3 }^{1 T_3}
\nonumber\\&\times
c_{\frac{\bm{P}}{2}-\bm{q}, s_z, \tau_3}
c_{\frac{\bm{P}}{2}+\bm{q}, s'_z, \tau'_3},\\
D_{1, S_z}(\bm{P}, \bm{q})
=\,&
\sum_{s_z, s'_z}
\sum_{\tau_3, \tau'_3}
C_{\frac{1}{2} \frac{1}{2} s_z s'_z}^{1 S_z}
C_{\frac{1}{2} \frac{1}{2}\tau_3\tau'_3 }^{00}
\nonumber\\&\times
c_{\frac{\bm{P}}{2}-\bm{q}, s_z, \tau_3}
c_{\frac{\bm{P}}{2}+\bm{q}, s'_z, \tau'_3}.
\end{align}
\end{subequations}
They can also be written explicitly as
\begin{subequations}\label{operatorP}
\begin{align}
P_{1, +1}(\bm{P}, \bm{q})
=\,&\frac{\sqrt{2}}{2}\left(
\nu_{\frac{\bm{P}}{2}-\bm{q}, \frac{1}{2}}
\nu_{\frac{\bm{P}}{2}+\bm{q}, -\frac{1}{2}}
-
\nu_{\frac{\bm{P}}{2}-\bm{q}, -\frac{1}{2}}
\nu_{\frac{\bm{P}}{2}+\bm{q}, \frac{1}{2}}
\right)
,\\
P_{1, 0}(\bm{P}, \bm{q})
=\,&\frac{1}{2}\left(
\nu_{\frac{\bm{P}}{2}-\bm{q}, \frac{1}{2}}
\pi_{\frac{\bm{P}}{2}+\bm{q}, -\frac{1}{2}}
-
\pi_{\frac{\bm{P}}{2}-\bm{q}, -\frac{1}{2}}
\nu_{\frac{\bm{P}}{2}+\bm{q}, \frac{1}{2}} \right.\nonumber\\
&\left.
+
\pi_{\frac{\bm{P}}{2}-\bm{q}, \frac{1}{2}}
\nu_{\frac{\bm{P}}{2}+\bm{q}, -\frac{1}{2}}
-
\nu_{\frac{\bm{P}}{2}-\bm{q}, -\frac{1}{2}}
\pi_{\frac{\bm{P}}{2}+\bm{q}, \frac{1}{2}}
\right)
,\\
P_{1, -1}(\bm{P}, \bm{q})
=\,&\frac{\sqrt{2}}{2}\left(
\pi_{\frac{\bm{P}}{2}-\bm{q}, \frac{1}{2}}
\pi_{\frac{\bm{P}}{2}+\bm{q}, -\frac{1}{2}}
-
\pi_{\frac{\bm{P}}{2}-\bm{q}, -\frac{1}{2}}
\pi_{\frac{\bm{P}}{2}+\bm{q}, \frac{1}{2}}
\right),
\end{align}
\end{subequations}
for the isovector channel and
\begin{subequations}\label{operatorD}
\begin{align}
D_{1, +1}(\bm{P}, \bm{q})
=\,&\frac{\sqrt{2}}{2}\left(
\nu_{\frac{\bm{P}}{2}-\bm{q}, \frac{1}{2}}
\pi_{\frac{\bm{P}}{2}+\bm{q}, \frac{1}{2}}
-
\pi_{\frac{\bm{P}}{2}-\bm{q}, \frac{1}{2}}
\nu_{\frac{\bm{P}}{2}+\bm{q}, \frac{1}{2}}
\right)
,\\
D_{1, 0}(\bm{P}, \bm{q})
=\,&\frac{1}{2}\left(
\nu_{\frac{\bm{P}}{2}-\bm{q}, \frac{1}{2}}
\pi_{\frac{\bm{P}}{2}+\bm{q}, -\frac{1}{2}}
-
\pi_{\frac{\bm{P}}{2}-\bm{q}, -\frac{1}{2}}
\nu_{\frac{\bm{P}}{2}+\bm{q}, \frac{1}{2}} \right.\nonumber\\
&\left.
-
\pi_{\frac{\bm{P}}{2}-\bm{q}, \frac{1}{2}}
\nu_{\frac{\bm{P}}{2}+\bm{q}, -\frac{1}{2}}
+
\nu_{\frac{\bm{P}}{2}-\bm{q}, -\frac{1}{2}}
\pi_{\frac{\bm{P}}{2}+\bm{q}, \frac{1}{2}}
\right)
,\\
D_{1, -1}(\bm{P}, \bm{q})
=\,&\frac{\sqrt{2}}{2}\left(
\nu_{\frac{\bm{P}}{2}-\bm{q}, -\frac{1}{2}}
\pi_{\frac{\bm{P}}{2}+\bm{q}, -\frac{1}{2}}
-
\pi_{\frac{\bm{P}}{2}-\bm{q}, -\frac{1}{2}}
\nu_{\frac{\bm{P}}{2}+\bm{q}, -\frac{1}{2}}
\right),
\end{align}
\end{subequations}
for the isoscalar channel.
The corresponding creation operators are their conjugates.

\begin{widetext}
Furthermore,using the symmetry of the interaction
\begin{align}
V_{t,s}(\bm{q}, \bm{q}')=V_{t,s}(\bm{q}, -\bm{q}')=V_{t,s}(-\bm{q}, \bm{q}')=V_{t,s}(-\bm{q}, -\bm{q}'),
\end{align}
the translational invariance of the system, and the operators shown in Eqs.~\eqref{operatorP} and \eqref{operatorD}, we obtain the explicit form of the Hamiltonian~\eqref{hamiltonian} as
\begin{align}
H=\,&\sum_{\bm{p}, s_z}\left(
\varepsilon_{\nu, \bm{p}}
\nu_{\bm{p}, s_z}^{\dagger} \nu_{\bm{p}, s_z}
+\varepsilon_{\pi, \bm{p}}
\pi_{\bm{p}, s_z}^{\dagger} \pi_{\bm{p}, s_z}\right)
\nonumber\\
&+\sum_{\bm{P}, \bm{q}, \bm{q}'}
V_s\left(\bm{q}, \bm{q}'\right) \left(
\nu_{\frac{\bm{P} }{ 2}+\bm{q}, \frac{1}{2}}^{\dagger}
\nu_{\frac{\bm{P} }{ 2}-\bm{q}, -\frac{1}{2}}^{\dagger}
\nu_{\frac{\bm{P} }{ 2}-\bm{q}', -\frac{1}{2}}
\nu_{\frac{\bm{P} }{ 2}+\bm{q}', \frac{1}{2}}
+
\pi_{\frac{\bm{P} }{ 2}+\bm{q}, \frac{1}{2}}^{\dagger}
\pi_{\frac{\bm{P} }{ 2}-\bm{q}, -\frac{1}{2}}^{\dagger}
\pi_{\frac{\bm{P} }{ 2}-\bm{q}', -\frac{1}{2}}
\pi_{\frac{\bm{P} }{ 2}+\bm{q}', \frac{1}{2}}
\right)
\nonumber\\
&+\sum_{\bm{P}, \bm{q}, \bm{q}'}\sum_{s_z}
V_t\left(\bm{q}, \bm{q}'\right)
\pi_{\frac{\bm{P} }{ 2}+\bm{q}, s_z}^{\dagger}
\nu_{\frac{\bm{P} }{ 2}-\bm{q}, s_z}^{\dagger}
\nu_{\frac{\bm{P} }{ 2}-\bm{q}', s_z}
\pi_{\frac{\bm{P} }{ 2}+\bm{q}', s_z}
\nonumber\\
&+\sum_{\bm{P}, \bm{q}, \bm{q}'}
\frac{V_s\left(\bm{q}, \bm{q}'\right)}{2}
\left(
\pi_{\frac{\bm{P} }{ 2}+\bm{q}, \frac{1}{2}}^{\dagger}
\nu_{\frac{\bm{P} }{ 2}-\bm{q}, -\frac{1}{2}}^{\dagger}
+\nu_{\frac{\bm{P} }{ 2}+\bm{q}, \frac{1}{2}}^{\dagger}
\pi_{\frac{\bm{P} }{ 2}-\bm{q}, -\frac{1}{2}}^{\dagger}
\right)
\left(
\nu_{\frac{\bm{P} }{ 2}-\bm{q}', -\frac{1}{2}}
\pi_{\frac{\bm{P} }{ 2}+\bm{q}', \frac{1}{2}}
+\pi_{\frac{\bm{P} }{ 2}-\bm{q}', -\frac{1}{2}}
\nu_{\frac{\bm{P} }{ 2}+\bm{q}', \frac{1}{2}}
\right)\nonumber\\
&+\sum_{\bm{P}, \bm{q}, \bm{q}'}
\frac{V_t\left(\bm{q}, \bm{q}'\right)}{2}
\left(
\pi_{\frac{\bm{P} }{ 2}+\bm{q}, \frac{1}{2}}^{\dagger}
\nu_{\frac{\bm{P} }{ 2}-\bm{q}, -\frac{1}{2}}^{\dagger}
-\nu_{\frac{\bm{P} }{ 2}+\bm{q}, \frac{1}{2}}^{\dagger}
\pi_{\frac{\bm{P} }{ 2}-\bm{q}, -\frac{1}{2}}^{\dagger}
\right)
\left(
\nu_{\frac{\bm{P} }{ 2}-\bm{q}', -\frac{1}{2}}
\pi_{\frac{\bm{P} }{ 2}+\bm{q}', \frac{1}{2}}
-\pi_{\frac{\bm{P} }{ 2}-\bm{q}', -\frac{1}{2}}
\nu_{\frac{\bm{P} }{ 2}+\bm{q}', \frac{1}{2}}
\right).
\end{align}
\end{widetext}

\subsection{Trial wave function}\label{sec:IIB}

Let us construct the trial wave function for quartet correlations in infinite symmetric nuclear matter.
Effects of the  center-of-mass  momentum $\bm{P}$ of a two-nucleon pair in the variational wave function has been shown to be nontrivial and beyond the mean-field framework, even for pure neutron matter or two-component Fermi gas along the BCS-BEC crossover~\cite{Ohashi}.
Consequently, in this study, $\bm{P}=0$ will be adopted for pair states along with the standard BCS theory as employed in the previous work~\cite{Senkov, Sandulescu, Sambataro, Sambataro2},
where pair states with time-conjugate orbitals are considered in finite nuclei.

By setting $\bm{P}=0$, the quartet creation operator can be introduced as
\begin{align}
\label{eq:alpha_creation}
\alpha^\dagger(\bm{q}, \bm{q}')
=\,&
\sum_{S_z, S'_z}
C_{11 S_z S'_z}^{00}
D_{1, S_z}^\dagger(0, \bm{q})D_{1, S'_z}^\dagger(0, \bm{q}')
\nonumber\\
=\,&
\frac{\sqrt{3}}{3}\left[
D_{1, +1}^\dagger(0, \bm{q})D_{1, -1}^\dagger(0, \bm{q}')\right.\nonumber\\&\left.
+D_{1, -1}^\dagger(0, \bm{q})D_{1, +1}^\dagger(0, \bm{q}')\right.\nonumber\\&\left.
-D_{1, 0}^\dagger(0, \bm{q})D_{1, 0}^\dagger(0, \bm{q}')\right].
\end{align}
Note that the corresponding quartet created by $\alpha^\dagger(\bm{q},\bm{q}')$ also has a zero center-of-mass momentum and hence forms the condensates in infinite matter.

For the four-body sector, the coherent state can be given by
\begin{align}
\left|\Psi_{\rm {coh }}\right\rangle
=\,&\exp \left(\sum_{\bm{q}, \bm{q}^{\prime}}
g_{\bm{q}, \bm{q}^{\prime}}
\alpha^{\dagger}
\left(\bm{q}, \bm{q}^{\prime}\right)\right)|0\rangle  \nonumber\\
=\,&\exp \left(g_{\bm{q}_{1}, \boldsymbol{q}_{1}} \alpha^{\dagger}\left(\bm{q}_{1}, \boldsymbol{q}_{1}\right)+g_{\boldsymbol{q}_{1}, \boldsymbol{q}_{2}} \alpha^{\dagger}\left(\boldsymbol{q}_{1}, \boldsymbol{q}_{2}\right)\right.\nonumber\\
&\left.
+g_{\boldsymbol{q}_{2}, \boldsymbol{q}_{1}} \alpha^{\dagger}\left(\boldsymbol{q}_{2}, \boldsymbol{q}_{1}\right)
+ \cdots\right)|0\rangle  \nonumber\\
=\,&\left(1+g_{\boldsymbol{q}_{1}, \boldsymbol{q}_{1}} \alpha^{\dagger}\left(\boldsymbol{q}_{1}, \boldsymbol{q}_{1}\right)\right)\left(1+g_{\boldsymbol{q}_{1}, \boldsymbol{q}_{2}} \alpha^{\dagger}\left(\boldsymbol{q}_{1}, \boldsymbol{q}_{2}\right)\right)\nonumber\\
&\times\left(1+g_{\boldsymbol{q}_{2}, \boldsymbol{q}_{1}} \alpha^{\dagger}\left(\boldsymbol{q}_{2}, \boldsymbol{q}_{1}\right)\right)\cdots|0\rangle \nonumber\\
=\,&\prod_{\boldsymbol{q}, \boldsymbol{q}^{\prime}}\left[1+g_{\boldsymbol{q}, \boldsymbol{q}^{\prime}} \alpha^{\dagger}\left(\boldsymbol{q}, \boldsymbol{q}^{\prime}\right)\right]|0\rangle,
\end{align}
with the weight parameter $g_{\bm{q},\bm{q}'}$.
This motivates us to use
\begin{align}
\left|\Phi\right\rangle=\,&
\prod_{\bm{q}, \bm{q}'}\left[
u_{\bm{q}, \bm{q}'}
+\sum_{S_z}
v_{\bm{q}, S_z} (\bm{q})
D_{1, S_z}^{\dagger}(0, \bm{q})\right.\nonumber\\&\left.
+\sum_{T_3}
 x_{\bm{q}, T_3} ({\bm{q}})
P_{1, T_3}^{\dagger}(0, \bm{q})
+w_{\bm{q}, \bm{q}'}
\alpha^{\dagger}(\bm{q}, \bm{q}')
\right]
|0\rangle,
\end{align}
as the trial wave function.
However, such a trial wave function is difficult to handle during the practical calculations even in a numerical way because of the multiple infinite products. For simplicity, the quartet creation operator is further approximated by taking the case of $\bm{q}=\bm{q}'$ as
\begin{align}
\alpha^\dagger(\bm{q})\equiv
\alpha^\dagger(\bm{q}, \bm{q}'=\bm{q}),
\end{align}
which describes a symmetric configuration with respect to relative momenta $\bm{q}$ of four nucleons.
The trial wave functional is then rewritten into
\begin{align}
\left|\Psi\right\rangle
=\,&\prod_{\bm{q}}
\left[
\sum_{T_3} x_{\bm{q}, T_3} P_{T_3}^{\dagger}(0, \bm{q})\right.\nonumber\\&\left.
+\left(\tilde u_{\bm{q}, +1}+\tilde v_{\bm{q}, +1} D_{1,+1}^{\dagger}(0, \bm{q})\right)
\right.\nonumber\\&\left.\times
\left(\tilde u_{\bm{q}, -1}+\tilde v_{\bm{q},-1} D_{1, -1}^{\dagger}({0}, \bm{q})\right)\right.\nonumber\\&\left.
+\left(\tilde u_{\bm{q}, 0}+\tilde v_{\bm{q}, 0} D_{1,0}^{\dagger}(0, \bm{q})\right)\right.\nonumber\\&\left.\times
\left(\tilde u'_{\bm{q}, 0}-\tilde v'_{\bm{q},0} D_{1, 0}^{\dagger}({0}, \bm{q})\right)
\right]
|0\rangle.
\end{align}
By expanding it, the trial wave function is finally obtained as
\begin{align}\label{twf1}
\left|\Psi\right\rangle
=\,&\prod_{\bm{q}}
\left[
u_{\bm{q}}
+\frac{\sqrt{2}}{2}\sum_{S_z}v_{\bm{q}, S_z} D_{1, S_z}^{\dagger}(0, \bm{q})\right.\nonumber\\&\left.
+\frac{\sqrt{2}}{2}\sum_{T_3} x_{\bm{q}, T_3} P_{T_3}^{\dagger}(0, \bm{q})
+\frac{1}{2}w_{\bm{q}, 1}\alpha^{\dagger}_1(\bm{q})\right.\nonumber\\&\left.
-\frac{1}{2}w_{\bm{q}, 2}\alpha^{\dagger}_2(\bm{q})
\right]
|0\rangle,
\end{align}
where
\begin{align}
\alpha_1^{\dagger}(\bm{q})= \,&D_{1,+1}^{\dagger}(0, \bm{q})D_{1, -1}^{\dagger}({0}, \bm{q}),\nonumber\\
\alpha_2^{\dagger}(\bm{q})= \,&D_{1,0}^{\dagger}(0, \bm{q})D_{1, 0}^{\dagger}({0}, \bm{q}),\nonumber\\
u_{\bm{q}}=\,&\tilde u_{\bm{q}, +1}\tilde u_{\bm{q}, -1},\quad
v_{\bm{q}, +1}=\tilde u_{\bm{q}, -1}\tilde v_{\bm{q}, +1},\nonumber\\
v_{\bm{q}, -1}=\,&\tilde u_{\bm{q}, +1}\tilde v_{\bm{q}, -1},\quad
v_{\bm{q}, 0}=\tilde u'_{\bm{q}, 0}\tilde v_{\bm{q}, 0}-\tilde u_{\bm{q}, 0}\tilde v'_{\bm{q}, 0},\nonumber\\
w_{\bm{q}, 1}=\,&\tilde v_{\bm{q}, +1}\tilde v_{\bm{q}, -1},\quad
w_{\bm{q}, 2}=\tilde v_{\bm{q}, 0}\tilde v'_{\bm{q}, 0}.
\end{align}
In addition, we define
\begin{align}
w_{\bm{q}}=w_{\bm{q}, 1}+w_{\bm{q}, 2}.
\end{align}
Consequently, the normalization condition reads
\begin{align}\label{norcon}
\sum_{S_z}\left|v_{\bm{q}, S_z}\right|^2
+\sum_{T_z}\left|x_{\bm{q}, T_3}\right|^2
+\left|u_{\bm{q}}\right|^2
+\left|w_{\bm{q}}\right|^2
=1.
\end{align}
It is remarkable that, in all of the following equations, only $w_{\bm{q}}$ appears, instead of any individual $w_{\bm{q}, 1}$ or $w_{\bm{q}, 2}$.
Due to this feature, an equivalent form of trial wave function to that in Eq.~\eqref{twf1} is the one without the term $\alpha_2^{\dagger}(\bm{q})= D_{1,0}^{\dagger}(0, \bm{q})D_{1, 0}^{\dagger}({0}, \bm{q})$, as the form taken in Ref.~\cite{Senkov}.
By introducing the ``lengths'' $\left|\bm{v}_{\bm{q}}\right|^{2}=\sum_{S_z}\left|v_{\bm{q}, S_z}\right|^{2}$ and $\left|\bm{x}_{\bm{q}}\right|^{2}=\sum_{T_3}\left|x_{\bm{q}, T_3}\right|^{2}$, we normalize the trial wave function according to
\begin{align}
\left|u_{\bm{q}}\right|^{2}+
\left|\bm{v}_{\bm{q}}\right|^{2}+
\left|\bm{x}_{\bm{q}}\right|^{2}+
\left|w_{\bm{q}}\right|^{2}
=1.
\end{align}

\subsection{Expectation value of Hamiltonian}\label{sec:IIC}

In this subsection, we will calculate the expectation value of Hamiltonian $\langle\Psi\left|H\right|\Psi\rangle$.

First, for the single-particle part, it is easy to obtain that
\begin{align}
&\langle\Psi\left|H_0\right|\Psi\rangle\nonumber\\
=\,&\langle\Psi\left|
\sum_{\bm{p}, s_z}
\left(\varepsilon_{\nu, \bm{p}}
\nu_{\bm{p}, s_z}^{\dagger} \nu_{\bm{p}, s_z}+\varepsilon_{\pi, \bm{p}}
\pi_{\bm{p}, s_z}^{\dagger} \pi_{\bm{p}, s_z}\right)
\right|\Psi\rangle\nonumber\\
=\,&\sum_{\bm{q}}\left[
\left|v_{\bm{q}, +1}\right|^2
\left(\varepsilon_{\pi, \bm{q}}+\varepsilon_{\nu, -\bm{q}}\right)
+\left|v_{\bm{q}, -1}\right|^2
\left(\varepsilon_{\pi, \bm{q}}+\varepsilon_{\nu, -\bm{q}}\right)\right.\nonumber\\
&\left.+\left|v_{\bm{q}, 0}\right|^2
\left(\varepsilon_{0, \bm{q}}+\varepsilon_{0, -\bm{q}}\right)
+\left|x_{\bm{q}, +1}\right|^2
\left(\varepsilon_{\nu, \bm{q}}+\varepsilon_{\nu, -\bm{q}}\right)
\right.\nonumber\\
&\left.
+\left|x_{\bm{q}, -1}\right|^2
\left(\varepsilon_{\pi, \bm{q}}+\varepsilon_{\pi, -\bm{q}}\right)
+\left|x_{\bm{q}, 0}\right|^2
\left(\varepsilon_{0, \bm{q}}+\varepsilon_{0, -\bm{q}}\right)
\right.\nonumber\\
&\left.
+2\left|w_{\bm{q}}\right|^2\left(
\varepsilon_{\pi, \bm{q}}+\varepsilon_{\nu, -\bm{q}}
\right)
\right],
\end{align}
where we have defined
\begin{align}
\varepsilon_{0, \bm{q}}=\frac{\varepsilon_{\pi, \bm{q}}+\varepsilon_{\nu, \bm{q}}}{2}.
\end{align}

At the next step, for the isoscalar pairing term, since it is orthogonal to the isovector part, i.e.,
\begin{align}
D_{1, S_z}(\bm{P}, \bm{q}')P^\dagger_{1, T_3}(0, \bm{q}_1)|0\rangle=0,
\end{align}
there is no overlap between them.
One can then obtain the contribution from the isoscalar part,
\begin{align}
&\langle\Psi\left|
\frac{1}{2} \sum_{\bm{P}, \bm{q}, \bm{q}'}
\sum_{S_z=-1}^{+1}
D_{1, S_z}^{\dagger}(\bm{P}, \bm{q})
V_t\left(\bm{q}, \bm{q}'\right)
D_{1, S_z}\left(\bm{P}, \bm{q}'\right)\right|\Psi\rangle\nonumber\\
=\,&
\sum_{\bm{q}, \bm{q}'}\sum_{S_z}
 V_t(\bm{q}, \bm{q}')
 \left[
 u_{\bm{q}}
 v_{\bm{q}, S_z}^{\ast}
+
\delta_{S_z, +1}
 v_{\bm{q}, -S_z}
 w_{\bm{q}}^{\ast}\right.\nonumber\\
&\left.+
\delta_{S_z, -1}
 v_{\bm{q}, -S_z}
 w_{\bm{q}}^{\ast}
 -
 \frac{1}{2}\delta_{S_z, 0}\left(
 v_{\bm{q}, -S_z}
 w_{\bm{q}}^{\ast}
 + v_{\bm{q}, -S_z}
 w_{-\bm{q}}^{\ast}
 \right)
 \right]\nonumber\\
 &\times
\left[
u_{\bm{q}'}^{\ast}
v_{\bm{q}', S_z}
+
\delta_{S_z, +1}
v_{{\bm{q}'}, -S_z}^{\ast}
w_{\bm{q}'}
+
\delta_{S_z, -1}
v_{{\bm{q}'}, -S_z}^{\ast}
w_{\bm{q}'}\right.\nonumber\\
&\left.-
\frac{1}{2}
\delta_{S_z, 0}
\left(
v_{{\bm{q}'}, -S_z}^{\ast}
w_{\bm{q}'}
+
v_{{\bm{q}'}, -S_z}^{\ast}
w_{-\bm{q}'}
\right)
\right],
\end{align}
where the terms with $\bm{q}=\bm{q}'$ which give the correction to one-body energies (i.e., the Hartree-Fock correction~\cite{FW}) are neglected, since they are smaller than the pairing terms with $\bm{q}\neq \bm{q}'$ which involve a large phase space associated with momenta for short-range attractive interactions.
Such an approximation is usually assumed in the BCS theory~\cite{FW,Schrieffer}.

\begin{widetext}
For the isovector part,
first obviously the isovector term is also orthogonal to the isoscalar one, which gives
\begin{align}
P_{1, T_3}(\bm{P}, \bm{q}')D^\dagger_{1, S_z}(0, \bm{q}_1)|0\rangle=0.
\end{align}
In a similar way, only pairing terms with $\bm{q}\neq \bm{q}'$ are considered here.
Consequently, it can be recast into
\begin{align}
&\langle\Psi\left|
\frac{1}{2} \sum_{\bm{P}, \bm{q}, \bm{q}'}
\sum_{T_3=-1}^{+1}
P_{1, T_3}^{\dagger}(\bm{P}, \bm{q})
V_s\left(\bm{q}, \bm{q}'\right)
P_{1, T_3}\left(\bm{P}, \bm{q}'\right)\right|\Psi\rangle\nonumber\\
=\,&
\frac{1}{2} \sum_{\bm{P}, \bm{q}, \bm{q}'}
\sum_{T_3=-1}^{+1}
\sum_{\tilde{\bm{q}}_1, \tilde{\bm{q}}_2}
\sum_{{\bm{q}}_1, {\bm{q}}_2}
V_s\left(\bm{q}, \bm{q}'\right)
\langle\psi({\tilde{\bm{q}}_2})\left|
P_{1, T_3}^{\dagger}(\bm{P}, \bm{q})
\right|\psi({\tilde{\bm{q}}_1})\rangle
\langle\psi({\bm{q}_2})\left|
P_{1, T_3}\left(\bm{P}, \bm{q}'\right)
\right|\psi({\bm{q}_1})\rangle,
\end{align}
where
\begin{align}
|\psi({\bm{k}})\rangle=\,&
\left[
u_{\bm{k}}
+\frac{\sqrt{2}}{2}\sum_{S_z}v_{\bm{k}, S_z} D_{1, S_z}^{\dagger}(0, \bm{k})
+\frac{\sqrt{2}}{2}\sum_{T_3} x_{\bm{k}, T_3} P_{T_3}^{\dagger}(0, \bm{k})
+\frac{1}{2}w_{\bm{k}}\alpha^{\dagger}(\bm{k})
\right]
|0\rangle.
\end{align}
As a result, one can obtain that the contribution from the isovector part as
\begin{align}
&\langle\Psi\left|
\frac{1}{2} \sum_{\bm{P}, \bm{q}, \bm{q}'}
\sum_{T_3=-1}^{+1}
P_{1, T_3}^{\dagger}(\bm{P}, \bm{q})
V_s\left(\bm{q}, \bm{q}'\right)
P_{1, T_3}\left(\bm{P}, \bm{q}'\right)\right|\Psi\rangle\nonumber\\
=\,&
\sum_{\bm{q}, \bm{q}'}\sum_{T_3}V_s\left(\bm{q}, \bm{q}'\right)
\left[x^\ast_{\bm{q}, T_3}u_{\bm{q}}
+\frac{1}{2}\delta_{T_3, 0}\left(
w^\ast_{\bm{q}}x_{\bm{q}, T_3}
+w^\ast_{-\bm{q}}x_{\bm{q}, T_3}\right)
+\delta_{{\bm{q}}, 0}\delta_{T_3, +1}w^\ast_{\bm{q}}x_{\bm{q}, T_3}
+\delta_{{\bm{q}}, 0}\delta_{T_3, -1}w^\ast_{\bm{q}}x_{\bm{q}, T_3}
\right]\nonumber\\
&\times\left[u^\ast_{\bm{q}'}x_{\bm{q}', T_3}
+\frac{1}{2}\delta_{T_3, 0}\left(
x^\ast_{\bm{q}', T_3}w_{\bm{q}'}
+x^\ast_{\bm{q}', T_3}w_{-\bm{q}'}\right)
+\delta_{{\bm{q}'}, 0}\delta_{T_3, +1}x^\ast_{\bm{q}', T_3}w_{\bm{q}'}
+\delta_{{\bm{q}'}, 0}\delta_{T_3, -1}x^\ast_{\bm{q}', T_3}w_{\bm{q}'}
\right].
\end{align}
Furthermore, the terms proportional to $\delta_{\bm{q}, 0}$ or $\delta_{\bm{q}', 0}$ are similar to those with $\bm{q} = \bm{q}'$, which give the correction to one-body energies, and thus they are neglected.
The above expression can then be rewritten into
\begin{align}
&\langle\Psi\left|
\frac{1}{2} \sum_{\bm{P}, \bm{q}, \bm{q}'}
\sum_{T_3=-1}^{+1}
P_{1, T_3}^{\dagger}(\bm{P}, \bm{q})
V_s\left(\bm{q}, \bm{q}'\right)
P_{1, T_3}\left(\bm{P}, \bm{q}'\right)\right|\Psi\rangle\nonumber\\
=\,&
\sum_{\bm{q}, \bm{q}'}\sum_{T_3}V_s\left(\bm{q}, \bm{q}'\right)
\left[x^\ast_{\bm{q}, T_3}u_{\bm{q}}
+\frac{1}{2}\delta_{T_3, 0}\left(
w^\ast_{\bm{q}}x_{\bm{q}, T_3}
+w^\ast_{-\bm{q}}x_{\bm{q}, T_3}\right)
\right]
\left[u^\ast_{\bm{q}'}x_{\bm{q}', T_3}
+\frac{1}{2}\delta_{T_3, 0}\left(
x^\ast_{\bm{q}', T_3}w_{\bm{q}'}
+x^\ast_{\bm{q}', T_3}w_{-\bm{q}'}\right)
\right].
\end{align}

By collecting all the terms, the expectation value of the Hamiltonian reads
\begin{align}
&\langle\Psi\left|
H
\right|\Psi\rangle\nonumber\\
=\,&\sum_{\bm{q}}\left[
\left|v_{\bm{q}, +1}\right|^2
\left(\varepsilon_{\pi, \bm{q}}+\varepsilon_{\nu, -\bm{q}}\right)
+\left|v_{\bm{q}, -1}\right|^2
\left(\varepsilon_{\pi, \bm{q}}+\varepsilon_{\nu, -\bm{q}}\right)
+\left|v_{\bm{q}, 0}\right|^2
\left(\varepsilon_{0, \bm{q}}+\varepsilon_{0, -\bm{q}}\right)
+\left|x_{\bm{q}, +1}\right|^2
\left(\varepsilon_{\nu, \bm{q}}+\varepsilon_{\nu, -\bm{q}}\right)\right.\nonumber\\&\left.
+\left|x_{\bm{q}, -1}\right|^2
\left(\varepsilon_{\pi, \bm{q}}+\varepsilon_{\pi, -\bm{q}}\right)
+\left|x_{\bm{q}, 0}\right|^2
\left(\varepsilon_{0, \bm{q}}+\varepsilon_{0, -\bm{q}}\right)
+2\left|w_{\bm{q}}\right|^2\left(
\varepsilon_{\pi, \bm{q}}+\varepsilon_{\nu, -\bm{q}}
\right)
\right]\nonumber\\
&+ \sum_{\bm{q}, \bm{q}'}\sum_{S_z}
 V_t(\bm{q}, \bm{q}')
 \left[
 u_{\bm{q}}
 v_{\bm{q}, S_z}^{\ast}
+
\delta_{S_z, +1}
 v_{\bm{q}, -S_z}
 w_{\bm{q}}^{\ast}
+
\delta_{S_z, -1}
 v_{\bm{q}, -S_z}
 w_{\bm{q}}^{\ast}
  -
  \frac{1}{2}\delta_{S_z, 0}\left(
 v_{\bm{q}, -S_z}
 w_{\bm{q}}^{\ast}
  +v_{\bm{q}, -S_z}
 w_{-\bm{q}}^{\ast}
 \right)
 \right]\nonumber\\
 &\times
\left[
u_{\bm{q}'}^{\ast}
v_{\bm{q}', S_z}
+
\delta_{S_z, +1}
v_{{\bm{q}'}, -S_z}^{\ast}
w_{\bm{q}'}
+
\delta_{S_z, -1}
v_{{\bm{q}'}, -S_z}^{\ast}
w_{\bm{q}'}
-
\frac{1}{2}\delta_{S_z, 0}\left(
v_{{\bm{q}'}, -S_z}^{\ast}
w_{\bm{q}'}
+
v_{{\bm{q}'}, -S_z}^{\ast}
w_{-\bm{q}'}
\right)
\right]\nonumber\\
&+\sum_{\bm{q}, \bm{q}'}\sum_{T_3}V_s\left(\bm{q}, \bm{q}'\right)
\left[x^\ast_{\bm{q}, T_3}u_{\bm{q}}
+\frac{1}{2}\delta_{T_3, 0}\left(
w^\ast_{\bm{q}}x_{\bm{q}, T_3}
+w^\ast_{-\bm{q}}x_{\bm{q}, T_3}\right)
\right]
\left[u^\ast_{\bm{q}'}x_{\bm{q}', T_3}
+\frac{1}{2}\delta_{T_3, 0}\left(
x^\ast_{\bm{q}', T_3}w_{\bm{q}'}
+x^\ast_{\bm{q}', T_3}w_{-\bm{q}'}\right)
\right].
\end{align}
\end{widetext}

\subsection{Variational equations}\label{sec:IID}

Since $u_{\bm{q}}$ can always be considered to be real, from the normalization condition~\eqref{norcon}, its variation with respect to $v^\ast_{\bm{q}, S_z}$, $x^\ast_{\bm{q}, T_3}$, and $w_{\bm{q}}^\ast$ can be expressed as
\begin{align}\label{veq1}
\delta u_{\bm{q}}=
-\frac{1}{2 u_{\bm{q}}}
\left(
\sum_{S_z} v_{\bm{q}, S_z}
\delta v_{\bm{q}, S_z}^{\ast}
+\sum_{T_3} x_{\bm{q}, T_3}
\delta x_{\bm{q}, T_3}^{\ast}
+w_{\bm{q}}
\delta w_{\bm{q}}^{\ast}\right).
\end{align}

\begin{widetext}
In analog to BCS, we can define the following energy gaps,
\begin{subequations}
\begin{align}\label{gapeq}
\Delta_{\bm{q}, S_z}^{\rm (isp)}&=
-\sum_{\bm{q}'}
V_t(\bm{q}, \bm{q}')
\left[
u_{\bm{q}'}^{\ast}
v_{\bm{q}', S_z}
+
\delta_{S_z, +1}
v_{\bm{q}', -S_z}^{\ast}
w_{\bm{q}'}
+
\delta_{S_z, -1}
v_{\bm{q}', -S_z}^{\ast}
w_{\bm{q}'}
-\frac{1}{2}\delta_{S_z, 0}\left(
v_{\bm{q}', -S_z}^{\ast}
w_{\bm{q}'}
+
v_{\bm{q}', -S_z}^{\ast}
w_{-\bm{q}'}
\right)
\right],\\
\Delta_{\bm{q}, T_3}^{\rm (ivp)}&=-\sum_{\bm{q}'}V_s\left(\bm{q}, \bm{q}'\right)
\left[u^\ast_{\bm{q}'}x_{\bm{q}', T_3}
+\frac{1}{2}\delta_{T_3, 0}\left(
x^\ast_{\bm{q}', T_3}w_{\bm{q}'}
+x^\ast_{\bm{q}', T_3}w_{-\bm{q}'}\right)
\right].
\end{align}
\end{subequations}

By performing the variations of the expectation value of Hamiltonian, one obtains
\begin{align}
\delta\langle\Psi\left|H\right|\Psi\rangle
=\,& v_{\bm{q}, +1}\delta v^\ast_{\bm{q}, +1}
\left(\varepsilon_{\pi, \bm{q}}+\varepsilon_{\nu, -\bm{q}}\right)
+v_{\bm{q}, -1}\delta v^\ast_{\bm{q}, -1}
\left(\varepsilon_{\pi, \bm{q}}+\varepsilon_{\nu, -\bm{q}}\right)
\nonumber\\&
+v_{\bm{q}, 0}\delta v^\ast_{\bm{q}, 0}
\left(\varepsilon_{0, \bm{q}}+\varepsilon_{0, -\bm{q}}\right)
+x_{\bm{q}, +1}\delta x^\ast_{\bm{q}, +1}
\left(\varepsilon_{\nu, \bm{q}}+\varepsilon_{\nu, -\bm{q}}\right)
\nonumber\\
&
+x_{\bm{q}, -1}\delta x^\ast_{\bm{q}, -1}
\left(\varepsilon_{\pi, \bm{q}}+\varepsilon_{\pi, -\bm{q}}\right)
+x_{\bm{q}, 0}\delta x^\ast_{\bm{q}, 0}
\left(\varepsilon_{0, \bm{q}}+\varepsilon_{0, -\bm{q}}\right)
\nonumber\\
&+2w_{\bm{q}}\delta w^\ast_{\bm{q}}
\left(\varepsilon_{\pi, \bm{q}}+\varepsilon_{\nu, -\bm{q}}\right)\nonumber\\
&- \sum_{S_z}
 \left[
v_{\bm{q}, S_z}^{\ast} \delta u_{\bm{q}}
 +u_{\bm{q}} \delta v_{\bm{q}, S_z}^{\ast}
+
\delta_{S_z, +1}
 v_{\bm{q}, -S_z}\delta w_{\bm{q}}^{\ast}\right.\nonumber\\&\left.
+
\delta_{S_z, -1}
 v_{\bm{q}, -S_z}\delta w_{\bm{q}}^{\ast}
  - \frac{1}{2}\delta_{S_z, 0}\left(
 v_{\bm{q}, -S_z}\delta w_{\bm{q}}^{\ast}
 + v_{-\bm{q}, -S_z}\delta w_{\bm{q}}^{\ast}
 \right)
 \right]\Delta_{\bm{q}, S_z}^{\rm (isp)}
\nonumber\\
&- \sum_{S_z}
\left[
v_{\bm{q}, -S_z}
\delta u_{\bm{q}}
+
\delta_{S_z, +1}
w_{\bm{q}}
\delta v_{\bm{q}, S_z}^{\ast}
+
\delta_{S_z, -1}
w_{\bm{q}}
\delta v_{\bm{q}, S_z}^{\ast}
-  \frac{1}{2}\delta_{S_z, 0}\left(
w_{\bm{q}}
\delta v_{\bm{q}, S_z}^{\ast}
+w_{-\bm{q}}
\delta v_{\bm{q}, S_z}^{\ast}
\right)
\right]{\Delta^\ast}_{\bm{q}, -S_z}^{\rm (isp)}\nonumber\\
&-\sum_{T_3}
\left[u_{\bm{q}}\delta x^\ast_{\bm{q}, T_3}
+x^\ast_{\bm{q}, T_3}\delta u_{\bm{q}}
+\frac{1}{2}\delta_{T_3, 0}x_{\bm{q}, T_3}\delta w^\ast_{\bm{q}}
+\frac{1}{2}\delta_{T_3, 0}x_{-\bm{q}, T_3}\delta w^\ast_{\bm{q}}
\right]
\Delta_{\bm{q}, T_3}^{\rm (ivp)}\nonumber\\
&- \sum_{T_3}
\left[x_{\bm{q}, T_3}\delta u_{\bm{q}}
+\frac{1}{2}\delta_{T_3, 0}w_{\bm{q}}\delta x^\ast_{\bm{q}, T_3}
+\frac{1}{2}\delta_{T_3, 0}w_{-\bm{q}}\delta x^\ast_{\bm{q}, T_3}
 \right]{\Delta^\ast}_{\bm{q}, T_3}^{\rm (ivp)},
\end{align}
where the symmetry
\begin{align}
\Delta_{\bm{q}, i}^{\rm j}=\Delta_{-\bm{q}, i}^{\rm j}
\end{align}
has been used.

Here we further introduce
\begin{align}
B_{\bm{q}}
=\,&
\frac{1}{2u_{\bm{q}}}\sum_{S_z, T_3}
\left[
v_{\bm{q}, S_z}^{\ast}\Delta_{\bm{q}, S_z}^{\rm (isp)}
+ v_{\bm{q}, S_z}{\Delta^\ast}_{\bm{q}, S_z}^{\rm (isp)}
+x_{\bm{q}, T_3}^{\ast}\Delta_{\bm{q}, T_3}^{\rm (ivp)}
+ x_{\bm{q}, T_3}{\Delta^\ast}_{\bm{q}, T_3}^{\rm (ivp)}\right]\nonumber\\
=\,&
\frac{1}{u_{\bm{q}}}{\rm Re}\left[
\bm{v}_{\bm{q}}^{\ast}\cdot\bm{\Delta}_{\bm{q}}^{\rm (isp)}
+\bm{x}_{\bm{q}}^{\ast}\cdot\bm{\Delta}_{\bm{q}}^{\rm (ivp)}
\right].
\end{align}
As a result, the variational equations can be obtained as
\begin{subequations}
\begin{align}
v_{\bm{q}, +1}=\,&\frac{
 u_{\bm{q}}\Delta_{\bm{q}, +1}^{\rm (isp)}
+ w_{\bm{q}}
{\Delta^\ast}_{\bm{q}, -1}^{\rm (isp)}
}{B_{\bm{q}}+\left(\varepsilon_{\pi, \bm{q}}+\varepsilon_{\nu, -\bm{q}}\right)},\quad
v_{\bm{q}, -1}=\frac{
 u_{\bm{q}}\Delta_{\bm{q}, -1}^{\rm (isp)}
+ w_{\bm{q}}
{\Delta^\ast}_{\bm{q}, +1}^{\rm (isp)}
}{B_{\bm{q}}+\left(\varepsilon_{\pi, \bm{q}}+\varepsilon_{\nu, -\bm{q}}\right)},\label{v1}\\
v_{\bm{q}, 0}=\,&\frac{
 u_{\bm{q}}\Delta_{\bm{q}, 0}^{\rm (isp)}
-
\frac{1}{2}\left( w_{\bm{q}}+ w_{-\bm{q}}\right)
{\Delta^\ast}_{\bm{q}, 0}^{\rm (isp)}
}{B_{\bm{q}}+\left(\varepsilon_{0, \bm{q}}+\varepsilon_{0, -\bm{q}}\right)},\label{v0}\\
x_{\bm{q}, +1}=\,&\frac{u_{\bm{q}}{\Delta}^{\rm (ivp)}_{\bm{q}, +1}
}{B_{\bm{q}}+\left(\varepsilon_{\nu, \bm{q}}+\varepsilon_{\nu, -\bm{q}}\right)},\quad
x_{\bm{q}, -1}=\frac{u_{\bm{q}}{\Delta}^{\rm (ivp)}_{\bm{q}, -1}}{B_{\bm{q}}+\left(\varepsilon_{\pi, \bm{q}}+\varepsilon_{\pi, -\bm{q}}\right)},\\
x_{\bm{q}, 0}=\,&\frac{u_{\bm{q}}{\Delta}^{\rm (ivp)}_{\bm{q}, 0}+\frac{1}{2}\left(w_{\bm{q}}+w_{-\bm{q}}\right){\Delta^\ast}^{\rm (ivp)}_{\bm{q}, 0}}{B_{\bm{q}}+\left(\varepsilon_{0, \bm{q}}+\varepsilon_{0, -\bm{q}}\right)},\\
w_{\bm{q}}=\,&\frac{
\frac{1}{2}\left(x_{\bm{q}, 0}+x_{-\bm{q}, 0}\right)
\Delta_{\bm{q}, 0}^{\rm (ivp)}
+ v_{\bm{q}, -1}\Delta_{\bm{q}, +1}^{\rm (isp)}
+ v_{\bm{q}, +1}\Delta_{\bm{q}, -1}^{\rm (isp)}
-\frac{1}{2}\left(v_{\bm{q}, 0}+v_{-\bm{q}, 0}\right)\Delta_{\bm{q}, 0}^{\rm (isp)}
 }{B_{\bm{q}}+2\left(\varepsilon_{\pi, \bm{q}}+\varepsilon_{\nu, -\bm{q}}\right)}.\label{eq:w_q}
\end{align}
\end{subequations}
\end{widetext}

\subsection{Number density equations and condensate fractions}\label{sec:IIE}

The number densities are given by
\begin{align}
n_\nu=\,&\sum_{\bm{q}}\left(\left|\bm{v}_{\bm{q}}\right|^2+2\left|x_{\bm{q}, +1}\right|^2+\left|x_{\bm{q}, 0}\right|^2+2\left|w_{\bm{q}}\right|^2\right)
\end{align}
for neutrons and
\begin{align}
n_\pi=\,&\sum_{\bm{q}}\left(\left|\bm{v}_{\bm{q}}\right|^2+2\left|x_{\bm{q}, -1}\right|^2+\left|x_{\bm{q}, 0}\right|^2+2\left|w_{\bm{q}}\right|^2\right)
\end{align}
for protons, respectively.
Thus the total density of nucleons is
\begin{align}
\label{eq:n_total}
n=n_\nu+n_\pi=\sum_{\bm{q}}\left(2\left|\bm{v}_{\bm{q}}\right|^2+2\left|\bm{x}_{\bm{q}}\right|^2+4\left|w_{\bm{q}}\right|^2\right).
 \end{align}

Being similar to the case of the conventional BCS theory~\cite{Salasnich}, the fractions of pairing and quartetting condensations can be calculated as
\begin{align}
f_{\rm pair}=\,&
\sum_{\bm{P},\bm{q}}
\left(\sum_{T_3}
\left|
\langle\Psi\left|
P_{1, T_3}\left(\bm{P}, \bm{q}\right)
\right|\Psi\rangle\right|^2\right.\nonumber\\
&\left.+
\sum_{S_z}
\left|\langle\Psi\left|
D_{1, S_z}\left(\bm{P}, \bm{q}\right)
\right|\Psi\rangle\right|^2\right)\nonumber\\
=\,&
\sum_{\bm{q}}\sum_{T_3}
\left|u^\ast_{\bm{q}}x_{\bm{q},T_3}+\delta_{T_3, 0}x^\ast_{\bm{q}, T_3}w_{\bm{q}}\right|^2\nonumber\\
&+
\sum_{\bm{q}}\sum_{S_z}
\left|u^\ast_{\bm{q}}v_{\bm{q},S_z}
-(-1)^{S_z}v^\ast_{\bm{q},S_z}w_{\bm{q}}\right|^2
,\\
f_{\rm quartet}=\,&
\sum_{\bm{q}}
\left|
\langle\Psi\left|
\alpha\left(\bm{q}\right)
\right|\Psi\rangle\right|^2
=\sum_{\bm{q}}
u_{\bm{q}}^2w_{\bm{q}}^2,
\end{align}
respectively.
Consequently, the condensate fractions depend on the variational parameters, which are obviously density dependent as one sees in Eq.~(\ref{eq:n_total}).
In addition, we also have applied the same approach for the electron-hole system to investigate the biexciton-like quartetting correlations~\cite{Guo}.
In this work, the density dependence of various physical quantities has been investigated, including the excitation gap (the minimum of energy dispersion), energy density, and chemical potential.
Deeper investigation on the condensation fraction with more realistic nuclear interactions will be useful, but it is out of scope of this paper.

\section{Results and discussion}\label{sec:III}

\subsection{Special case of $w_{\bm{q}}= 0$}

First, let us discuss the special case without the quartet correlations ($w_{\bm{q}}= 0$) to see how the well-known BCS pairing state is realized in our framework.
In such a special case, the isoscalar and isovector pairings play similar roles, and thus we take the isoscalar pairing as an example.
Correspondingly, from Eqs.~\eqref{v1} and \eqref{v0} we obtain
\begin{align}
v_{\boldsymbol{q}, S_{z}}=\frac{u_{\boldsymbol{q}} \Delta_{\boldsymbol{q}, S_{z}}^{\text {(isp) }}}{B_{\boldsymbol{q}}+2 \varepsilon_{\boldsymbol{q}}},
\end{align}
where we take $\varepsilon_{\boldsymbol{q}, \nu}=\varepsilon_{\boldsymbol{q}, \pi}=\varepsilon_{\boldsymbol{q}}$ and assume that the variational parameters are real for simplicity.
We then obtain
\begin{align}\label{BqBCS}
B_{\boldsymbol{q}}=\frac{\sum_{S_{z}}\left|\Delta_{\boldsymbol{q}, S_{z}}^{(\mathrm{isp})}\right|^{2}}{B_{\boldsymbol{q}}+2 \varepsilon_{\boldsymbol{q}}}= \frac{\bm{\Delta}_{\boldsymbol{q}}^{2}}{B_{\boldsymbol{q}}+2 \varepsilon_{\boldsymbol{q}}},
\end{align}
and thus
\begin{align}
B_{\bm{q}}^{2}+2 \varepsilon_{\bm{q}} B_{\bm{q}}-\bm{\Delta}_{\bm{q}}^{2}=0,
\end{align}
i.e.,
\begin{align}
B_{\bm{q}}
= -\varepsilon_{\bm{q}} \pm \sqrt{\varepsilon_{\bm{q}}^{2}+\bm{\Delta}_{\bm{q}}^{2}}.
\end{align}
When we choose
$B_{\boldsymbol{q}}=-\varepsilon_{\boldsymbol{q}}+\sqrt{\varepsilon_{\boldsymbol{q}}^{2}+\bm{\Delta}_{\boldsymbol{q}}^{2}}$
(noting that this ambiguity of the sign originates from the physical broken U(1) symmetry associated with the phase of the gap parameter), we get
\begin{align}
v_{\boldsymbol{q}, S_{z}}=\frac{\Delta_{\boldsymbol{q}, S_{z}}^{(\mathrm{isp})}}{\sqrt{\varepsilon_{\boldsymbol{q}}^{2}+\bm{\Delta}_{\boldsymbol{q}}^{2}}+\varepsilon_{\boldsymbol{q}}} u_{\boldsymbol{q}},
\end{align}
and we can define
\begin{align}\label{E1}
E=\sqrt{\varepsilon_{\boldsymbol{q}}^{2}+\bm{\Delta}_{\boldsymbol{q}}^{2}}.
\end{align}
By using the normalization condition, we can further obtain
\begin{subequations}
\begin{align}
u_{\bm{q}}^{2}=\,&\frac{1}{2}\left(1+\frac{\varepsilon_{\bm{q}}}{\sqrt{\varepsilon_{\bm{q}}^{2}+\bm{\Delta}_{\bm{q}}^{2}}}\right), \\
\quad \bm{v}_{\bm{q}}^{2}=\,&\frac{1}{2}\left(1-\frac{\varepsilon_{\bm{q}}}{\sqrt{\varepsilon_{\bm{q}}^{2}+\bm{\Delta}_{\bm{q}}^{2}}}\right).\label{vq}
\end{align}
\end{subequations}
These are the well-known results of BCS theory.

Because of the spin degeneracy ${\bm{\Delta}_{\bm{q}}}^{2}=\left|\Delta_{\bm{q}, -1}^{(\text {isp })}\right|^{2}+\left|\Delta_{\bm{q}, 0}^{(\text {isp })}\right|^{2}+\left|\Delta_{\bm{q}, +1}^{(\mathrm{isp})}\right|^{2}=3\left|\Delta_{\bm{q}}^{\text {(isp) }}\right|^{2}$, the gap equation~\eqref{gapeq} reads
\begin{align}\label{gapnew}
\Delta_{\bm{q}, S_{z}}^{(\mathrm{isp})}
=\,&-\sum_{\bm{q}^{\prime}} V_{t}\left(\boldsymbol{q}, \boldsymbol{q}^{\prime}\right) \frac{\Delta_{\bm{q}', S_{z}}^{(\mathrm{isp})} }{2 \sqrt{\varepsilon_{\boldsymbol{q}^{\prime}}^{2}+\bm{\Delta}_{\bm{q}^{\prime}}^{2}}} \nonumber\\
=\,&-\sum_{\boldsymbol{q}^{\prime}} V_{t}\left(\boldsymbol{q}, \boldsymbol{q}^{\prime}\right) \frac{ \Delta_{\bm{q}^{\prime}}^{(\mathrm{isp})}}{2 \sqrt{\varepsilon_{\boldsymbol{q}^{\prime}}^{2}+\bm{\Delta}_{\boldsymbol{q}^{\prime}}^{2}}}.
\end{align}
Moreover, if the interactions are separable, for example, $V_{t}\left(\boldsymbol{q}, \boldsymbol{q}^{\prime}\right)=-\lambda_{t} e^{-\bm{q}^{2} / b^{2}} e^{-\bm{q}^{\prime 2} / b^{2}}$ with the coupling constant $\lambda_t$ and the range parameter $b$~\cite{Ropke}, we can write the gap as $\Delta_{\bm{q}}^{(\text {isp })}=\phi e^{-\bm{q}^{2} / b^{2}}$ with a constant $\phi$. Finally, the gap equation reads
\begin{align}
1= \lambda_{t} \sum_{\boldsymbol{q}^{\prime}} e^{-2 \bm{q}^{\prime 2} / b^{2}} \frac{1}{2 \sqrt{\varepsilon_{\boldsymbol{q}^{\prime}}^{2}+\bm{\Delta}_{\bm{q}^{\prime}}^{2}}}.
\end{align}
Once $\phi$ is determined by solving the above equation, thermodynamic quantities such as the internal energy and number density can be obtained, together with $u_{\bm{q}}$ and $v_{\bm{q}, S_{z}}$.
In this way, one can find that the momentum dependence of the gap parameters is associated with that of the interactions. Note that we obtain the momentum-independent gap parameters for the contact-type (i.e., momentum-independent) couplings.

\subsection{Quartet correlations}

When the quartet correlations are taken in account, in a similar way to Eq.~\eqref{BqBCS}, one can obtain
\begin{align}
B_{\bm{q}} &=\frac{\sum_{S_{z}}\left|\Delta_{\bm{q}, S_{z}}^{(\mathrm{isp})}\right|^{2}+\frac{w_{\bm{q}}}{u_{\bm{q}}}\left[2 \Delta_{\bm{q},+1}^{(\mathrm{isp})} \Delta_{\bm{q},-1}^{(\mathrm{isp})}-\left|\Delta_{\bm{q}, 0}^{(\mathrm{isp})}\right|^{2}\right]}{B_{\bm{q}}+2 \varepsilon_{\bm{q}}}\nonumber \\
&=\frac{\Delta_{\bm{q}}^{2}+R_{\bm{q}}}{B_{\bm{q}}+2 \varepsilon_{\bm{q}}},
\end{align}
where $R_{\bm{q}}$ is defined as
\begin{align}
R_{\bm{q}}=\frac{w_{\bm{q}}}{u_{\bm{q}}}\left[2 \Delta_{\bm{q},+1}^{(\mathrm{isp})} \Delta_{\bm{q},-1}^{(\mathrm{isp})}-\left|\Delta_{\bm{q}, 0}^{(\mathrm{isp})}\right|^{2}\right].
\end{align}
One can then obtain
\begin{align}
B_{\bm{q}}=-\varepsilon_{\bm{q}}+\sqrt{\varepsilon_{\bm{q}}^{2}+\bm{\Delta}_{\bm{q}}^{2}+R_{\bm{q}}},
\end{align}
and recast the energy dispersion~\eqref{E1} into
\begin{align}\label{E2}
E=\sqrt{\varepsilon_{\bm{q}}^{2}+\bm{\Delta}_{\bm{q}}^{2}+R_{\bm{q}}}.
\end{align}
In addition, one can obtain
\begin{align}
w_{\bm{q}}=\,&\frac{u_{\bm{q}}\left[2 \Delta_{\bm{q},+1}^{(\mathrm{isp})} \Delta_{\bm{q},-1}^{(\mathrm{isp})}-\left|\Delta_{q, 0}^{(\mathrm{isp})}\right|^{2}\right]+w_{\bm{q}} \sum_{S_{z}}\left|\Delta_{\bm{q}, S_{z}}^{(\mathrm{isp })}\right|^{2}}{\left(B_{\bm{q}}+4 \varepsilon_{\bm{q}}\right)\left(B_{\bm{q}}+2 \varepsilon_{\bm{q}}\right)}\label{wcorr}
\end{align}
and
\begin{align}
\frac{w_{\bm{q}}}{u_{\bm{q}}}=\,&\frac{\left[2 \Delta_{\bm{q},+1}^{(\mathrm{isp})} \Delta_{\bm{q},-1}^{(\mathrm{isp})}-\left|\Delta_{\bm{q}, 0}^{(\mathrm{isp})}\right|^{2}\right]}{\left(B_{\bm{q}}+4 \varepsilon_{\bm{q}}\right)\left(B_{\bm{q}}+2 \varepsilon_{\bm{q}}\right)-\sum_{S_{z}}\left|\Delta_{\bm{q}, S_{z}}^{(\mathrm{isp})}\right|^{2}}.
\end{align}
Note that the contribution of the isovector pairing is ignored in the above equations as in the previous subsection.
Nevertheless, this approximation is still valid for qualitative understanding of in-medium quartet correlations in infinite nuclear matter since the isovector pairing is much weaker than the isoscalar pairing~\cite{Tajima}.

\begin{figure}
  \includegraphics[width=0.45\textwidth]{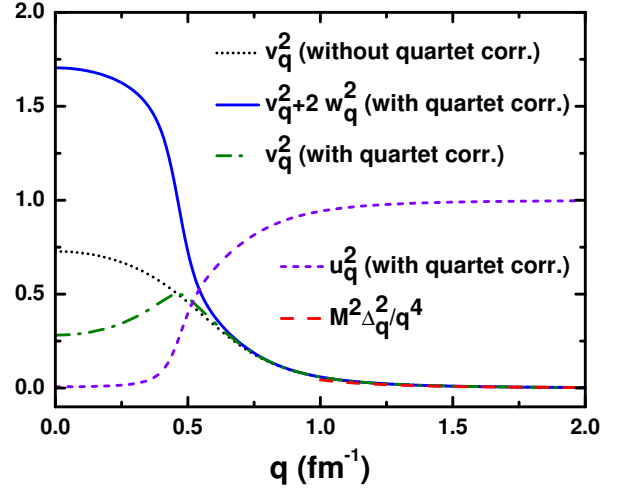}
  \caption{(Color online) Solutions to Eqs.~\eqref{vq} and \eqref{wcorr} as a function of relative momentum $\bm{q}$.
  Here $\Delta^{\rm (isp)}_{\bm{q}, S_z}$ and $\Delta^{\rm (isp)}_{\bm{q}, S_z}/\mu$ are adopted as $5$~MeV and $1.16$ ($\mu=4.31$~MeV)~\cite{TajimaPRX}, respectively.
  The results $\bm{v}^2_{\bm{q}}+2w^2_{\bm{q}}$, $\bm{v}^2_{\bm{q}}$, and $u_{\bm{q}}^2$ to Eq.~\eqref{wcorr} with the quartet correlation are shown with the blue solid, olive dash-dotted, and violet short-dashed lines, respectively.
  The result $\bm{v}^2_{\bm{q}}$ to Eq.~\eqref{vq} without the quartet correlation is shown with the black short-dotted line.
  The curve $M^2\bm{\Delta}^2_{\bm{q}}/q^4$ is also shown with the red dashed line for the asymptotic behavior.}\label{fig:1}
\end{figure}

In order to investigate the asymptotic behavior of the variational parameters, the interaction is taken as a contact-type one.
By taking the energy gap $\Delta^{\rm (isp)}_{\bm{q}, S_z}$ to be $5$~MeV (noting that $|\bm{\Delta}_{\bm{q}}|$ should be multiplied by another factor $\sqrt{3}$) and the ratio $\Delta^{\rm (isp)}_{\bm{q}, S_z}/\mu=1.16$~\cite{TajimaPRX} as a typical value of Fermi superfluids, the solutions to Eqs.~\eqref{vq} and \eqref{wcorr} are plotted in Fig.~\ref{fig:1}.
The results $\bm{v}^2_{\bm{q}}+2w^2_{\bm{q}}$, $\bm{v}^2_{\bm{q}}$, and $u_{\bm{q}}^2$ to Eq.~\eqref{wcorr} with the quartet correlation as a function of relative momentum $\bm{q}$ are shown with the blue solid, olive dash-dotted lines, and violet short-dashed, respectively.
In contrast, the result $\bm{v}^2_{\bm{q}}$ to Eq.~\eqref{vq} without the quartet correlation is shown with the black short-dotted line.
It can be seen that with the consideration of quartet correlation, nucleons prefer to form the more stable $\alpha$-like particles, in the low-momentum region.
With the increase of relative momentum $\bm{q}$, $\alpha$-like particles break into pairs, and thus $w_{\bm{q}}^2$ gets smaller while $\bm{v}_{\bm{q}}^2$ gets larger.
When the relative momentum $\bm{q}$ is high enough to even break up the pairs, both $w_{\bm{q}}^2$ and $\bm{v}_{\bm{q}}^2$ decrease.
The curve $M^2\bm{\Delta}^2_{\bm{q}}/q^4$ is also shown with the red dashed line for reference of the high-momentum tail.
This result is consistent with the so-called Tan's relation~\cite{Tan1,Tan2,Tan3} for the BCS pairing, which is widely discussed in cold-atom physics and also in nuclear systems in terms of short-range correlations~\cite{Hen,Weiss,Urban}.
Physically, the power-law tail $M^2\bm{\Delta}^2_{\bm{q}}/q^4$ describes high-momentum nucleons forming neutron-proton Cooper pairs due to the short-range attraction.
Although the quartet correlation does not contribute to the high-momentum tail in our framework,
such an effect may appear via the correlations associated with excited deuterons with finite center-of-mass momenta, which are neglected in this study and will be addressed elsewhere.

\begin{figure}[t]
  \includegraphics[width=0.45\textwidth]{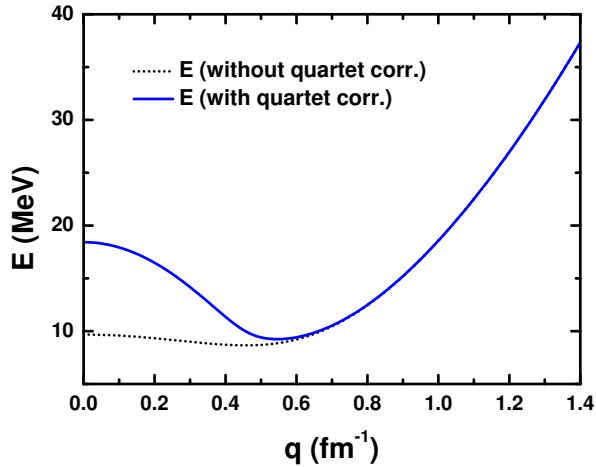}
  \caption{(Color online) Energy dispersion $E$ with and without the quartet correlation as a function of relative momentum $\bm{q}$. The energy dispersion without the quartet correlation~\eqref{E1} is shown with the black short-dotted line, and the one with correlation~\eqref{E2} is shown with the blue solid line.
  The same parameters in Fig.~\ref{fig:1} are adopted.}\label{fig:2}
\end{figure}

Still taking $\Delta^{\rm (isp)}_{\bm{q}, S_z}=5$~MeV and $\Delta^{\rm (isp)}_{\bm{q}, S_z}/\mu=1.16$, the energy dispersions with and without the quartet correlation are shown in Fig.~\ref{fig:2} as a function of relative momentum $\bm{q}$.
The correction to the energy dispersion caused by the quartet correlation mainly contribute to the low relative momentum region.
When the relative momentum become larger than the value which gives the minimum of the energy dispersion, the quartet correlation becomes almost negligible.
It is because that the excited deuterons are not considered here.
It can be seen that the value of the excitation gap also becomes larger.
This behavior is reasonable in the sense that a larger energy is required to cause a single-nucleon excitation at low momenta accompanying the breakups of not only a deuteron-like Cooper pair but also an $\alpha$-like Cooper quartet.

Furthermore, it should be noted that quartet correlations in the present approach vanish in the absence of the pairing gaps as found in Eq.~(\ref{eq:w_q}).
This behavior is associated with the assumption on the quartet creation operator $\alpha^\dagger(\bm{q},\bm{q}')$ in the variational wave function, which consists of the combinations of two isoscalar pair creation operators with zero center-of-mass momentum $\bm{P}=\bm{0}$ [see also Eq.~(\ref{eq:alpha_creation})].
As a consequence, the quartet correlations always involve only condensed pairs (i.e., $\bm{P}=\bm{0}$) in the present approach.
Nevertheless, it can be assumed that the fraction of condensed pairs is sufficiently larger than that of uncondensed ones with $\bm{P}\neq\bm{0}$ at low temperature.
To generalize the wave functions for the case with quartet correlations but without condensed pairs,
one may consider the quartet creation operator $\tilde{\alpha}^\dag(\bm{q},\bm{q}',\bm{P})$ involving excited pairs as
\begin{align}
\tilde{\alpha}^\dagger(\bm{q}, \bm{q}',\bm{P})
=\,&
\sum_{S_z, S'_z}
C_{11 S_z S'_z}^{00}
D_{1, S_z}^\dagger(\bm{P}, \bm{q})D_{1, S'_z}^\dagger(-\bm{P}, \bm{q}'),
\end{align}
where zero center-of-mass momentum of the four-body state is kept [i.e., $\bm{P}+(-\bm{P})=\bm{0}$].
However, as we mentioned in Sec.~\ref{sec:IIB},
the trial wave function associated the coherent state of
$\tilde{\alpha}^\dag(\bm{q},\bm{q}',\bm{P})$ with multiple momentum degrees of freedom is practically difficult to handle in the numerical computation.
A similar approximation was employed in the previous work for finite nuclei~\cite{Senkov,Sandulescu,Baran,Baran2}, considering pairing states with the time-conjugate orbitals.
More explicitly, $\tilde{\alpha}^\dag(\bm{q}, \bm{q}',\bm{P})$ can be rewritten in the form of
\begin{align}
  \tilde{\alpha}^\dagger(\bm{q}, \bm{q}',\bm{P})
  =\alpha^\dag(\bm{q})\delta_{\bm{P},\bm{0}}\delta_{\bm{q},\bm{q}'}+\delta\alpha^\dag(\bm{q},\bm{q}',\bm{P}),
\end{align}
where $\delta\alpha^\dag(\bm{q},\bm{q}',\bm{P})\propto (1-\delta_{\bm{P},\bm{0}})$ is the operator for the contribution associated with excited pairs, which is neglected in this study.

\begin{figure}
  \includegraphics[width=0.45\textwidth]{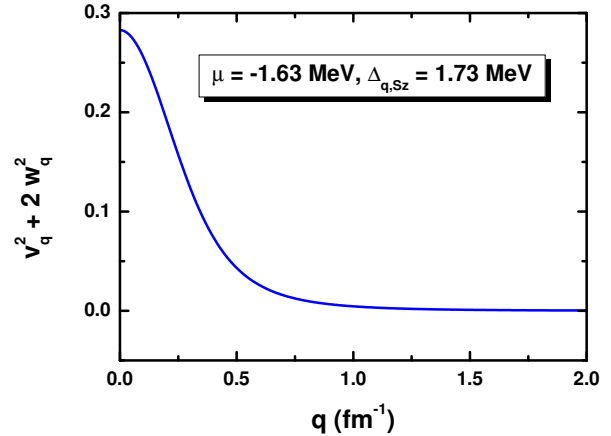}
  \caption{(Color online) Solution to Eq.~\eqref{wcorr} as a function of relative momentum $\bm{q}$ in the low-density regime (i.e., negative chemical potential $\mu=-1.63$~MeV~\cite{Sogo2010}).
  Here $\Delta^{\rm (isp)}_{\bm{q}, S_z}$ is adopted as $1.73$~MeV ($|\bm{\Delta}_{\bm{q}}|=3$~MeV), which is comparable with the previous numerical result of the neutron-proton pairing gap in symmetric nuclear matter~\cite{Lombardo}.
  }\label{fig:3}
\end{figure}

As we mentioned that the high-momentum component of the dispersion in Fig.~\ref{fig:2} may be affected by excited pairs,
such a tendency can also be found in the variational parameters at the low-density regime where $\mu$ is negative and the $\alpha$-particle formation is dominant.
For that, the solution to Eq.~\eqref{wcorr} as a function of relative momentum $\bm{q}$ in the low density regime is shown in Fig.~\ref{fig:3}.
In order to qualitatively compare it with the occupation number $\rho(k)$ of nucleons ($k$ is a nucleon momentum) calculated by solving the in-medium four-body Schr\"{o}dinger equation in Ref.~\cite{Sogo2010}, the chemical potential $\mu$ is taken as $-1.63$~MeV.
Although the specific value of $\bm{\Delta}_{\bm{q}}$ was not shown in Ref.~\cite{Sogo2010}, here $\Delta^{\rm (isp)}_{\bm{q}, S_z}$ is adopted as $1.73$~MeV (i.e., $|\bm{\Delta}_{\bm{q}}|=3$~MeV), which is comparable with the previous numerical result of the neutron-proton pairing gap in symmetric nuclear matter~\cite{Lombardo}.
It is found that $v_{\bm{q}}^2+2w_{\bm{q}}^2$ approaches almost zero around $q= 1.0$~fm$^{-1}$ in the present calculation,
while $\rho(k)$ in Ref.~\cite{Sogo2010} remains a relatively non-negligible finite value even above $k=1.0$ fm$^{-1}$ (noting that the overall absolute value depends on details of the configurations).
Although the calculated quantities, the approaches, and the configurations (e.g., interactions) are different from each other,
the quickly decreasing behavior of $v_{\bm{q}}^2+2w_{\bm{q}}^2$ with increasing the relative momentum $\bm{q}$ may be reminiscent of the lack of excited pair contributions in the present approach. 

As a step further, we may also consider the effective four-body interaction
\begin{align}
H_4=\sum_{\bm{q},\bm{q}'}V_4(\bm{q},\bm{q}')\alpha^\dag(\bm{q})\alpha(\bm{q}'),
\end{align}
which has been taken into account in the study for finite nuclei~\cite{Senkov}.
In such a case, one can obtain an additional gap-like parameter $\Delta_{\bm{q}}^{(\alpha)}$ representing solely the quartet correlations without pairing gaps.
In other words, the quartet component will not vanish even if the pairing condensates are absent, in the presence of four-body interaction terms in the Hamiltonian.
Although the coupling strength $V_4(\bm{q},\bm{q}')$ is still unknown in infinite nuclear matter, in principle it should be nonzero.
Even for infinitesimally small $V_4(\bm{q},\bm{q}')$,
such a multibody interaction can affect in-medium quartet properties, as it has been found that the three- and four-body interactions are crucial for the formation of in-medium fermionic clusters~\cite{Akagami,Tajima2,Nishida}.
The inclusion of these multibody interactions are left for future work.

\section{Summary and Perspectives}\label{sec:IV}

We theoretically investigated the $\alpha$-like quartet correlations in the cold nuclear matter at the thermodynamic limit.
The BCS-type variational wave function in the momentum space is extended to the infinite symmetric nuclear matter, where the pair and quartet correlations are described by the coherent superposition of neutron-proton Cooper pairs and $\alpha$-like Cooper quartets.
We showed an analytic structure of the quartet BCS wave function in terms of nucleon degrees of freedom
and the hierarchical structure of in-medium cluster formations in the momentum space in the cold symmetric nuclear matter.

In particular, we examined the quasiparticle dispersion and the momentum distribution of nucleons under the presence of deuteron-like pair and $\alpha$-like quartet condensations.
With a low relative momentum $\bm{q}$ (and zero center-of-mass momentum), nucleons prefer to form the $\alpha$-like Cooper quartet, which is more stable {than the deuteron-like Cooper pairs}.
With the increase of relative momentum $\bm{q}$, $\alpha$-like Cooper quartets break into two Cooper pairs, and the {fraction} of {deuteron-like} pairs increases monotonically.
When the relative momentum $\bm{q}$ is high enough to even kinematically break up the pair, the fractions of both pairs and quartets decrease with exhibiting the high-momentum-tail behavior associated with residual pairs.
The quartet correlations also give a significant correction to the energy dispersion in the low-momentum region.
The low-energy excitation of nucleons involves a larger energy gap compared to the usual BCS pairing state because it requires the breakups of Cooper pairs as well as Cooper quartets in the nuclear matter.
The present results may also contribute to the interdisciplinary understanding of multicomponent fermionic condensations beyond the BCS paradigm in many-body systems, and improve the understanding of many-body correlations in the nuclear system.

In the next step, by further taking realistic forms of interactions and gradually upgrading the wave function, one can investigate the quartet correlations in a deeper way.
By using the variational equations, one can calculate different kinds of equations of states (EOS) for the nuclear matter to investigate various properties, such as the saturation properties.
Furthermore, while we assumed that the order parameters are real valued in the present study, it would be interesting to explore possible intrinsic Josephson currents and out-of-phase collective modes in the quartet BCS framework with complex order parameters.

\begin{acknowledgments}

The authors are grateful to Tomoya Naito for the fruitful discussions.
Y.G. was supported by RIKEN Junior Research Associate Program.
H.T. acknowledges the JSPS Grants-in-Aid for Scientific Research under Grant No.~18H05406.
H.L. acknowledges the JSPS Grant-in-Aid for Early-Career Scientists under Grant No.~18K13549, the JSPS Grant-in-Aid for Scientific Research (S) under Grant No.~20H05648, and the RIKEN Pioneering Project: Evolution of Matter in the Universe.
\end{acknowledgments}

\end{CJK}

\end{document}